\documentclass[]{article}

\usepackage[noblocks]{authblk}
\usepackage{amsmath}
\usepackage{amssymb}
\usepackage{graphicx}
\usepackage{hyperref}
\usepackage{xcolor}
\usepackage[latin1]{inputenc}

\newcommand{\sbk}[1]{\left[ #1 \right]}
\newcommand{\rbk}[1]{\left( #1 \right)}
\newcommand{\cbk}[1]{\left\{ #1 \right\}}

\DeclareMathOperator{\erfc}{erfc}
\DeclareMathOperator{\sign}{sign}

\begin{document}

\title{Generalization from correlated sets of patterns in the perceptron}

\author[1]{Francesco Borra}
\author[2,3]{Marco Cosentino Lagomarsino}
\author[4,5]{Pietro Rotondo}
\author[2,6]{Marco Gherardi}

\affil[1]{Universit\`a degli studi di Roma La Sapienza, Italy}
\affil[2]{Universit\`a degli studi di Milano, via Celoria 16, Milano, Italy}
\affil[3]{IFOM Foundation FIRC Institute of Molecular Oncology, Milan Italy}
\affil[4]{School of Physics and Astronomy, University of Nottingham, Nottingham, NG7 2RD, UK}
\affil[5]{Centre for the Mathematics and Theoretical Physics of Quantum Non-equilibrium Systems, University of Nottingham, Nottingham NG7 2RD, UK}
\affil[6]{INFN Sezione di Milano}

\maketitle

\begin{abstract}
  Generalization is a central aspect of learning theory.
  Here, we propose a framework that explores an auxiliary \emph{task-dependent}
  notion of generalization, and attempts to quantitatively
  answer the following question: given two sets of patterns with a
  given degree of dissimilarity, how easily will a network be able to
  ``unify'' their interpretation? This is quantified by the volume of
  the configurations of synaptic weights that classify the two sets in
  a similar manner. To show the applicability of our idea in a
  concrete setting, we compute this quantity for the perceptron, a
  simple binary classifier, using the classical statistical physics
  approach in the replica-symmetric ansatz.
  In this case, we show how an analytical expression measures the
  ``distance-based capacity'', the maximum load of patterns
  sustainable by the network, at fixed dissimilarity between patterns
  and fixed allowed number of errors.  This curve indicates that
  generalization is possible at any distance, but with decreasing
  capacity.
  We propose that a distance-based definition of generalization may be
  useful in numerical experiments with real-world neural networks, and
  to explore computationally sub-dominant sets of synaptic solutions.
\end{abstract}

%
%

\section{Introduction}

Generalization is an essential feature of cognition.  It constructs
broad, universal statements or general concepts from a few empirical
observations. Our ability to generalize comprises a wide and not well
characterized set of tasks and abilities, including the tendency to
respond in the same way to different, but similar ``on some level'',
stimuli.  An important feature of generalization is to assume or
recognize the existence of common features shared by sets of
elements. Hence, classifying common relations and differences among
different observed patterns is crucial~\cite{Gluck:2011}.

The problem of breaking down the process of generalization into its
elementary components naturally emerges in the field of artificial
neural networks.  In this context, achieving a deeper understanding of
generalization could improve the current design principles and
clarify the reasons why current architectures generalize well.
A generally held belief is that the efficiency of deep neural networks
in identifying patterns can be understood in terms of feature
extraction.  Despite its robustness, this view is being challenged by an increasing
number of experimental results.  For example, a recent study showed
that this paradigm is misleading when trying to understand how deep
neural networks generalize, and proposes to consider alternative
approaches~\cite{Zhang:arXiv:2016}; one such approach already emerged
in a statistical mechanics
setting~\cite{Martin:arXiv:2017,Seung:PRA:1992}.  Another challenging
observation is a recently discovered fragility of neural networks to
so-called ``adversarial attacks''~\cite{Goodfellow:arXiv:2013},
whereby any given pattern can be maliciously modified into a very
similar pattern that gets misclassified.

Learning efficiently from examples requires a way to gauge the size
and quality of the training and test sets.
The classic methods to address these issues (e.g., bias-variance
decomposition, Vapnik-Chervonenkis dimension, model complexity) are
found within so-called ``statistical learning
theory''~\cite{Vapnik:IEEE:1999}. In this framework, the goal of
learning is the approximation of a function $f$ by an element $h$ of a
given hypothesis space $\mathcal H$.  The trained machine $h$ is
chosen by a learning algorithm, starting from a training set, i.e.~a
set of pairs $\xi_\mu,f(\xi_\mu)$, where $\xi_\mu$ are chosen from a
space $\Omega$ with unknown probability distribution.
Statistical learning theory establishes general relations between the
complexity or expressivity of a given function class $\mathcal H$ and
its generalization properties, defined as the ability of its elements
to reproduce the output of $f$ beyond the training
set~\cite{Mehta:arXiv:2018}.  Despite its power, this framework has
limitations concerning its applicability to neural
networks~\cite{Friedland:arXiv:2018}.  For instance, it was observed
recently that over-parameterized neural networks often generalize
better than smaller, less complex, ones.  This finding is in conflict
with the predictions of statistical learning and classical
computational complexity theory (as well as with naive
intuition)~\cite{Novak:arXiv:2018}.

Another important drawback of statistical learning theory is that it
considers generalization in a worst-case scenario, where the
capabilities of the network are tested against possibly malicious
counterexamples~\cite{Opper:2001}.
A statistical physics approach overcomes this problem by considering
the so-called ``teacher-student'' scenario
\cite{Seung:PRA:1992,Monasson:PRL:1995,Levin:IEEE:1990}.  This
framework usually assumes that the generalization ability of a trained
network (the student) is measured on a set of tests produced by a
teacher with identical architecture.
An important limitation of the teacher-student setting is that the
student can learn from irrelevant examples if the teacher is used as a
surrogate for a classification rule not involving the whole space of inputs.  Consider, for instance, a teacher trained to discern
handwritten digits from a training set $\Omega_\mathrm{t}$ (e.g. from
the popular MNIST database).  The training set is a subset of a much
larger set $\Omega$ of all black-and-white images.  The teacher will
associate well-defined output values to elements of
$\Omega_\mathrm{t}$ but also to any other element of $\Omega$, for
instance to those in a set $\Omega_\mathrm{n}$, disjoint from
$\Omega_\mathrm{t}$, containing random noise.  Now the student can in
principle learn to reproduce the teacher's response on
$\Omega_\mathrm{t}$ by learning to mimic the teacher on the
(meaningless) set $\Omega_\mathrm{n}$.  In other words, the student
can learn to classify a dataset without ever seeing a single example
from it.
This problem is connected to the more general fact that the space
$\Omega$, in practical applications, has structure, first and foremost
via notions of distance or similarity between its elements.
%

Building on these premises, we propose here an auxiliary way to
approach generalization. The main object of interest is the number of
different synaptic configurations that allow the network to classify
correctly two sets of patterns. This quantity is a generalization of
the classic ``Gardner
volume''~\cite{Gardner:JPA:1988,Theumann:JPA:1991} to the case where
the training set consists of two sets of patterns with a prescribed
degree of correlation.
The rationale is to take explicitly into account the degree of
similarity between ``new'' and ``old'' information seen by a network.
Here we use a simple definition of distance between binary data sets,
based on their overlaps, but the framework applies to any other
definition of distance.    
The load at which the Gardner volume becomes zero defines a capacity,
which in our case depends on the distance.
To root our ideas in a concrete case, we carry out these calculations
on a very simple neural network, the perceptron, using classic results
from statistical mechanics~\cite{Gardner:JPA:1988, Nishimori:2001}.


\section{Definition of the approach and results}

\subsection{Measuring the similarity between classification tasks}

\label{similaritysection}
This section defines two complementary notions of dissimilarity
between classification tasks, one based on a bit-wise comparison of
the training sets, and the other based on the free-energy landscape of
the corresponding supervised learning problem.

A classifier neural network is described by an output function
$\sigma$ that associates an output $\sigma(\boldsymbol{\xi}_\mu)$ to
any input vector $\boldsymbol{\xi}_\mu\in \mathbb{R}^N$. We consider
binary input vectors, whose elements are $\pm 1$ bits and we define a
classification task as a training set
$\xi=\{\boldsymbol{\xi}_\mu\}_{\mu=1,...,p}$ with associated labels
$\sigma=\{\sigma_\mu\}_{\mu=1,...,p}$. Therefore, a task is a set of
input-output pairs
$(\xi,\sigma)=\{(\boldsymbol{\xi}_\mu,\sigma_\mu)\}_{\mu=1,..,.p}$. The
network can solve the task if $\forall\mu=1,...,p$,
$\sigma_\mu=\sigma(\boldsymbol{\xi}_\mu)$.

Let us fix two classification tasks, say $(\xi,\sigma)$ and
$(\bar\xi,\bar\sigma)$.  We focus on inputs ($\xi$ and $\bar\xi$) and
outputs ($\sigma$ and $\bar\sigma$) separately, thus defining two
bit-wise distances.
The canonical distance between two patterns $\boldsymbol{\xi}_\mu$ and
$\boldsymbol{\xi}_\nu$ is the normalized Hamming distance, defined as
\begin{equation}
  d_{\textrm{H}}(\boldsymbol{\xi}_\mu,\boldsymbol{\xi}_\nu)=\frac{1}{2}-
  \frac{1}{2N} \boldsymbol{\xi}_\mu\cdot\boldsymbol{\xi}_\nu. 
\end{equation}
The quantity
$\boldsymbol{\xi}_\mu\cdot\boldsymbol{\xi}_\nu/N = \sum_{i=1}^N
\xi_\mu^i\xi_\nu^i/N =: q(\boldsymbol{\xi}_\mu,\boldsymbol{\xi}_\nu)$
is the overlap between $\boldsymbol{\xi}_\mu$ and
$\boldsymbol{\xi}_\nu$.  We need to extend this definition to the full
sets of inputs $\xi=\{\boldsymbol{\xi}_\mu\}_{\mu=1,\ldots,p}$ and
$\bar\xi=\{\bar{\boldsymbol{\xi}}_{\bar\mu}\}_{\bar\mu=1,\ldots, p}$.
Intuitively, $\xi$ and $\bar\xi$ are identical if, for any
$\boldsymbol{\xi}_\mu\in\xi$ there exists some
$\bar{\boldsymbol{\xi}}_{\bar\mu}\in\bar\xi$ such that
$\boldsymbol{\xi}_\mu=\bar{\boldsymbol{\xi}}_{\bar\mu}$.  Following
this line of reasoning, one can define a distance between the sets
$\xi$ and $\bar\xi$ as
\begin{equation}
  \label{d}
  d_{\textrm{in}}(\xi,\bar\xi)=\min_{\pi\in
    S_p}\sum_{\mu}d(\boldsymbol{\xi}_\mu,\bar{\boldsymbol{\xi}}_{\pi(\mu)}), 
\end{equation}
where $S_p$ is the symmetric group of order $p$.  The minimum over
$S_p$ selects the closest matching between each pattern in $\xi$ and
each one in $\bar\xi$, thus restoring the symmetry under permutations,
but it complicates the analytical computations.  Therefore, we adopt a
simpler setup, still inspired by (\ref{d}), where the elements of
the input sets are at fixed distance $d$ pairwise, meaning that
$d_{\textrm{in}}(\xi_\mu,\bar\xi_\mu)=d_\textrm{in}$ for every $\mu$. 
We consider the ensemble of $p$ independent pairs of inputs at Hamming
distance $d_{\textrm{in}}$, defined by the joint probability
\begin{align}
P_{d_{\textrm{in}}}(\xi,\bar\xi):&=P(\xi,\bar\xi|d_{\textrm{in}}(\xi,\bar\xi)=d_{\textrm{in}})
  \\ 
&= \frac{1}{P(d_\textrm{in})^p}\prod_{\mu=1}^p
        P(\boldsymbol{\xi}_\mu)P(\bar{\boldsymbol{\xi}}_\mu)\,\delta(d_{\textrm{H}}(\boldsymbol{\xi}_\mu,\bar{\boldsymbol{\xi}}_\mu)-d_{\textrm{in}}) 
\end{align}
where
$P(d_\textrm{in}):=P(d_{\textrm{H}}(\boldsymbol{\xi}_1,\boldsymbol{\xi}_2)=d_{\textrm{in}})$
is the so-called Bayesian normalization, i.e., the probability that a
pair of two random vectors $\boldsymbol{\xi}_1$ and
$\boldsymbol{\xi}_2$ has Hamming distance $d_{\textrm{in}}$.

Having fixed the matching between inputs by this definition, it is
then natural to define the distance between outputs
$\sigma=\{\sigma_\mu\}_{\mu=1,...,p}$ and
$\bar\sigma=\{\bar\sigma_{\bar\mu}\}_{\bar\mu=1,...,p}$ as
\begin{equation}
d_{\textrm{out}}(\sigma,\bar\sigma)=\frac{1}{p}\sum_{\mu=1}^p
(1-\delta_{\sigma_\mu,\bar\sigma_{\mu}}). 
\end{equation}

We now focus on the similarity between two tasks in terms of synaptic
representation.  We reason that, from the point of view of the network,
two tasks are similar if they are easily solved by the same synaptic
configuration, i.e., if they share a significant fraction of common
solutions.
Here, by solutions we mean the solutions to the problem of finding a
state of the network, specified by a synaptic weight structure $W$,
such that the task is solved correctly.
Two equivalent tasks will share all solutions and should have zero
synaptic distance.  Conversely, two tasks are incompatible if they
have no common solution, or, equivalently, if there exists no solution
to the task of learning the union of the two tasks.
In this section, we do not specify the nature of $W$ any further,
since our approach can be applied to more general multi-layered
architectures. In the following sections we apply our ideas to the
case of the perceptron only.

A definition of synaptic distance consistent with these intuitive
requirements can be formalized as follows.  Let us consider a neural
network with cost function $H_{\xi,\sigma}(W)$, equal, for instance,
to the number of errors that the network makes in performing the task
$(\xi,\sigma)$ when its synaptic weights are $W$.  Let the cost
function be additive under union of the training sets: given two sets
of input-output pairs $(\xi,\sigma)$ and $(\bar\xi,\bar\sigma)$ of
arbitrary sizes $p$ and $\bar p$,
\begin{equation}
H_{(\xi,\sigma)\cup(\bar\xi,\bar\sigma)}(W)=H_{\xi,\sigma}(W)+H_{\bar\xi,\bar\sigma}(W),
\end{equation}
where $(\xi,\sigma)\cup(\bar\xi,\bar\sigma)$ denotes the labelled
training set with inputs
$\left\{\boldsymbol{\xi}_1,\ldots,\boldsymbol{\xi}_p,\bar{\boldsymbol{\xi}}_1,\ldots,\bar{\boldsymbol{\xi}}_{\bar
    p}\right\}$ and outputs
$\left\{\sigma_1,\ldots,\sigma_p,\bar\sigma_1,\ldots,\bar\sigma_{\bar
    p}\right\}$.
The canonical partition function at inverse temperature $\beta$ of a
system with Hamiltonian $H_{\xi,\sigma}$ is
\begin{equation}\label{gardnervolume}
Z(\beta,(\xi,\sigma))=\int d W \;e^{-\beta H_{\xi,\sigma}(W)},
\end{equation}
where $dW$ is a normalized measure on the synaptic weights.
In the zero-temperature limit $\beta\to\infty$, the system occupies
the lowest-energy state and $Z$ becomes the degeneracy of the ground
state, or the fraction of exact solutions to the task
$(\xi,\sigma)$. The free energy is, in this context, referred to as
the Gardner volume:
\begin{equation}\label{gv}F(\beta,(\xi,\sigma)) = -\frac{1}{\beta N}\ln\,Z(\beta,(\xi,\sigma))\end{equation}
Our definition of network distance is then
\begin{equation}\label{dm}
\Omega_{\textrm{n}}((\xi,\sigma),(\bar\xi,\bar\sigma),\beta):=
\frac{1}{\beta N}
\ln \frac{\sqrt{ Z(\beta,(\xi,\sigma)) Z(\beta,(\bar\xi,\bar\sigma)) } }{Z(\beta,(\xi,\sigma)\cup(\bar\xi,\bar\sigma))}.
\end{equation}
In the zero-temperature limit $\beta\to\infty$ this quantity
counts the number of exact common solutions to the two tasks,
normalized by the number of solutions to the two tasks separately.
In fact, Eq.~(\ref{dm}) can be rewritten in terms of the free energy
as
\begin{equation}\label{dm2}
\Omega_\textrm{n}((\xi,\sigma),(\bar\xi,\bar\sigma),\beta)=-\frac{F(\xi,\sigma)+F(\bar\xi,\bar\sigma)}{2}+F((\xi,\sigma)\cup (\bar\xi,\bar\sigma))
\end{equation}
The synaptic distance is zero for two identical tasks and diverges for
incompatible tasks:
$\Omega_\textrm{n}((\xi,\sigma),(\xi,\sigma),\infty)=0$ and
$\Omega_\textrm{n}((\xi,\sigma),(\xi,-\sigma),\infty)=\infty$.

In this setting, the ability of the network to generalize can be
studied by comparing the two types of distances defined above.  In
particular, we will consider the typical value of $\Omega$ in the
ensemble where $d_{\textrm{in}}=d_{\textrm{in}}(\xi,\bar\xi)$ and
$D_{\textrm{out}}=d_{\textrm{out}}(\sigma,\bar\sigma)$ are fixed:

\begin{equation}\label{avd}
{\Omega} (D_{\textrm{out}},d_{\textrm{in}})=\left\langle
  \Omega_\textrm{n}((\xi,\sigma),(\bar\xi,\bar\sigma),\beta)\right\rangle_{d_{\textrm{in}}(\xi,\bar\xi)=d_{\textrm{in}};\;\;
  d_{\textrm{out}}(\sigma,\bar\sigma)=D_{\textrm{out}}}. 
\end{equation}
A critical line is identified by the point in which $\Omega=\infty$
for fixed $(D_{\textrm{out}},d_{\textrm{in}})$ for a given size of the
input sets. Equivalently, for fixed $d_{\textrm{in}}$ and size $p$,
$\Omega=\infty$ indicates the theshold values of $D_{\textrm{out}}$,
i.e. the range of output-similarity that the network can typically
attribute to the input sets.

%
%

\subsection{Generalization properties of the perceptron}
\label{model}

We now set out to specify the abstract notion of memory-based distance
introduced in the previous section (Eq.~\eqref{avd}) in order to use
it for an explicit calculation. We call
$\xi=\{\boldsymbol{\xi}_\mu\}_{\mu=1,...,p}$ a set of $p=\alpha N$
input vectors with components $\xi_\mu^i=\pm1\;\;\forall
i=1,...,N$. The perceptron is a network which yields a binary output
$\sigma_W(\xi_\mu)=\sign(\boldsymbol{W}\cdot
\boldsymbol{\xi}_\mu/N-K)=\pm 1$ for any input, given a certain
synaptic configuration $\boldsymbol{W}\in\mathbb R^N$. In the
spherical model $\sum_i W_i^2=N$, in the discrete model $W_i=\pm1$.

In the case of batch learning, for any given training set
$(\xi,\sigma)$, the energy conventionally associated to this model is
the error-counting cost function
\begin{equation}
H_{\xi,\sigma}(\boldsymbol{W})=\sum_{\mu=1}^p\Theta(-\sigma_{\mu}\;\;\boldsymbol{\xi}_\mu\cdot \boldsymbol{W}).
\end{equation}
This definition allows to compute the Gardner volume,
i.e.~the fraction of synaptic configurations $\boldsymbol{W}$ that
solve a certain task, as defined in (\ref{gv}).

As for the probabily of synaptic configurations $dP(\boldsymbol{W})$,
the following maximally entropic probability distributions are
conventionally used, for the spherical and discrete case respectively
\begin{equation}
\begin{split}
dP(\boldsymbol{W})&=\frac{d \boldsymbol{ W}\;\delta\left( \sum_{i=1}^N
W_i^2 -N\right)}{\int_{\mathbb{R}^N} d
\boldsymbol{W}\;\delta\left( \sum_{i=1}^N W_i^2 -N\right)} \quad \mathrm{(spherical)}\\
dP(\boldsymbol{W})&=\frac{1}{2^N} \quad \mathrm{(discrete)}.
\end{split}
\end{equation}
A different cost function was proposed recently, in order to study a
peculiar clustering property of the solutions in the perceptron with
discrete weigths~\cite{Huang:PRE:2014,Baldassi:PRL:2015,Baldassi:PNAS:2016}.

In order to study the typical behaviour of this network, the Gardner
volume should be averaged on the input-output pair statistics
$P(\xi,\sigma)$ (quenched average). A second-order phase transition
is witnessed by the average value of the cost function. The
critical capacity $\alpha_c(\beta)$ is defined as
\begin{equation}
  \label{ac}\alpha_c(\beta)=\inf_{\alpha}\{\alpha:\;\left\langle
    H\right\rangle_{(\xi,\sigma)} >0\}
\end{equation}
and identifies a critical line in the $(\alpha,\beta)$
plane. Physically, $\alpha_c(\beta)$ is the maximum number of patterns
per neuron that can be learned in the typical case (without committing
an extensive number of mistakes). The value of $\alpha_c$ clearly
depends on the statistics of the input-ouput pairs. The simplest
statistics is obtained by choosing both inputs and outputs randomly
and independently
\begin{equation}
  \label{random} 
  P(\xi,\sigma)=\prod_{\mu=1}^p
  [a\,\delta_{\sigma_\mu,-1}+(1-a)\,\delta_{\sigma_\mu,1}]\prod_{j=1}^N
  [b\,\delta_{\xi_\mu^j,-1}+(1-b)\,\delta_{\xi^j_\mu,1}]  \ .
\end{equation}

In the unbiased case, $a=b=1/2$ and $\beta=\infty$, $\alpha_c=2$ for
the spherical perceptron, while $\alpha_c\approx0.833$ in the discrete
case. It is important to point out that, in the latter case, the
replica-symmetric ansatz yields a quantitatively incorrect result and
one-step replica symmetry breaking is needed \cite{Theumann:JPA:1993}.

In the case of a single set of independent and spatially correlated
inputs, two classic studies~\cite{Monasson:JP:1993,Monasson:JPA:1992},
derive the capacity from the (intra) correlation matrix
$C_{ij}=\langle\xi_\mu^i \xi_\mu^j\rangle$. More recent studies
\cite{Kabashima:JP:2008,Kabashima:JPA:2009} have focused on the
capacity in the case of a prescribed correlation between different
patterns, as given by the overlap matrix
$C_{\mu\nu}=\boldsymbol{\xi}_\mu\cdot\boldsymbol{\xi}_\nu/N$.

The teacher-student setting for generalization assumes the following
specific choice of the output statistics. Instead of drawing inputs
and outputs independently, there is an input (example) statistics
$P(\xi)$ and a teacher machine, described by an output function
$\sigma_T$. For each input, the ``correct'' output is chosen by the
teacher. Specifically, the input-output statistics is given by
\begin{equation} \label{teacherstudent}
  P(\xi,\sigma)=P(\xi)\prod_{\mu=1}^p\delta(\sigma_\mu-\sigma_T(\xi_\mu)) 
\end{equation} 
in a noise-free scenario.  In this scenario, the cost function (called
``learning function''), is algorithm-specific, and its average value
$\langle H\rangle$, as a function of the number of examples
$\alpha=p/N$, is called ``learning curve'' $\epsilon(\alpha)$.  The
learning curve quantifies generalization. The generalization ability
can be defined by averaging on all possible teachers, if needed. For a
perceptron, the best possible performance is achieved with the
Bayesian algorithm which, however, can only be performed by a perceptron-based committee
machine~\cite{Kinouchi:PRE:1996,Opper:PRL:1991,Nishimori:2001}.

\subsubsection{Distance-based Gardner volume}

Turning to our approach to define generalization, we choose both
inputs $\xi$ and outputs $\sigma$ to be random and unbiased. Since we
have two sets $(\xi,\sigma)$ and $(\bar\xi,\bar\sigma)$, the standard
gauge freedom of the problem allows to fix the outputs of the first
set $\sigma_\mu$ to $+1$.
With this premise, we write the non-trivial part of Eq.~(\ref{dm2}),
for given $d_{\mathrm{out}}(\sigma,\bar\sigma)=D_{\mathrm{out}}$, as
\begin{align}\label{ourvolume}
F(D_{\mathrm{out}})=\frac{1}{N}&\ln\int dP(\boldsymbol{W})\;\\
&\sum_{\{\epsilon_\mu=\pm 1\}}\delta\Big(\sum_{\mu=1}^p\epsilon_\mu-p(1-2D_{\textrm{out}})\Big)\;\prod_{\mu=1}^p\Theta(\boldsymbol{W}\cdot \boldsymbol{\xi}_\mu)\;\Theta(\epsilon_\mu \boldsymbol{W}\cdot\bar{\boldsymbol{\xi}}_{\bar\mu}).
\end{align}
Note that this is a ``zero-temperature'' definition, and that
\begin{displaymath}
  \sum_{\{\epsilon_\mu=\pm
    1\}}\delta\Big(\sum_{\mu=1}^p\epsilon_\mu-p(1-2D_{\textrm{out}})\Big)\;\prod_{\mu=1}^p\Theta(\boldsymbol{W}\cdot
  \boldsymbol{\xi}_\mu)\;\Theta(\epsilon_\mu
  \boldsymbol{W}\cdot\bar{\boldsymbol{\xi}}_{\bar\mu}) 
\end{displaymath}
equals one if exactly $D_{\textrm{out}}p$ output pairs are discordant,
regardless of which pairs, and zero otherwise. This choice of summing
over $\epsilon_\mu$ inside the logarithm makes Eq.~(\ref{ourvolume})
slightly different from Eq.~(\ref{dm2}). 
Nonetheless, the two quantities are equal
up to a combinatorial prefactor which is irrelevant for the final result. 
%
This also justifies the conditional notation
$(D_{\textrm{out}}|d_{\textrm{in}})$ once the average over input sets
of given distance is taken.

The full expression for (\ref{avd}) is therefore
\begin{equation}\Omega(D_{\textrm{out}}|d_{\textrm{in}})=\overline{
    F_\alpha}-\overline{
    F_\alpha(D_{\textrm{out}}|d_{\textrm{in}})}\end{equation}
where $\overline{ F_{\alpha}}=\overline{ F_{\alpha}(0|0)}$ is the
conventional Gardner volume for $N\alpha$ inputs, and  
\begin{equation}\label{f}
\overline{ F_\alpha(D_{\textrm{out}}|d_{\textrm{in}})}=\int P_{d_{\textrm{in}}}(\xi,\bar\xi) F_{\alpha}(\xi,\bar\xi, D_{\textrm{out}})\,.
\end{equation} 
Both these observables can be evaluated within the replica formalism,
as will be outlined in the next section (a detailed derivation will be
given in Appendix \ref{A} and \ref{B}).

Finally, we introduce the distance-based capacity
$\alpha_c(D_{\textrm{out}}|d_{\textrm{in}})$, which is to
$\overline{F_\alpha(D_{\textrm{out}}|d_{\textrm{in}})}$ what
$\alpha_c$ (Eq.~(\ref{ac})) is to the Gardner volume
(\ref{gardnervolume}). Since in our ``zero temperature'' framework we
have not explicitly introduced a cost function, it
is convenient to define $\alpha_c(D_{\textrm{out}}|d_{\textrm{in}})$ as
\begin{equation}\label{ouralpha}
\alpha_c(D_{\textrm{out}}|d_{\textrm{in}})=\inf_\alpha\{\alpha: \overline{F_\alpha(D_{\textrm{out}}|d_{\textrm{in}})}>-\infty\}.
\end{equation}
Physically, $\alpha_c(D_{\textrm{out}}|d_{\textrm{in}}) N$ is the
maximum size of two sets, with distance $d_{\textrm{in}}$, that can be
learned simultaneously with
$D_{\textrm{out}}\; \alpha_c(D_{\textrm{out}}|d_{\textrm{in}}) N$
concordant output pairs.

Now, suppose the perceptron has learned a set $\xi$ of size
$\alpha N$. If a new set $\bar\xi$, with
$d_{\textrm{in}}(\xi,\bar\xi)=d_{\textrm{in}}$ is presented, then
there will be a fraction $D_{\textrm{out}}$ of discordant outputs and
in this case $\alpha_c(D_{\textrm{out}}|d_{\textrm{in}})$ shows the
range $D_{\textrm{out}}$ can assume, given $d_{\textrm{in}}$. It must
be remarked that this boundary does not show whether any
$D_{\textrm{out}}$ has a finite probability. A related quantity is the
conditional probability $P(D_{\textrm{out}}|d_{\textrm{in}})$, which
may be approximated by
\begin{align}
P(D_{\textrm{out}}|d_{\textrm{in}})&\approx\frac{\exp(N F(D_{\textrm{out}}|d_{\textrm{in}}))}{\int dD'\;\exp(N F(D'|d_{\textrm{in}}))}\\
&=\exp[N F(D_{\textrm{out}}|d_{\textrm{in}})-N\max_{D'}F(D'|d_{\textrm{in}})]\to\left\{\begin{matrix}1&\mbox{finite chance}\\ 0&\mbox{no chance}\end{matrix}\right.
\end{align}
However, this idea requires some non-trivial
computations and careful geometrical considerations, and we do not pursue it here.

\subsubsection{Analytical expression for the distance-based capacity}

We are able to derive an analytical expression for the critical
capacity at fixed pattern distance, which we discuss in the following.
The quantity $\overline{F_{\alpha}(D_{\textrm{out}}|d_{\textrm{in}})}$
in Eq.~(\ref{f}) is, formally, an average free energy and can be computed
with the replica formalism \cite{Gardner:JPA:1988,Nishimori:2001}. 
The replica method is based on the identity
\begin{equation}
\overline{ F_\alpha(D_{\textrm{out}}|d_{\textrm{in}})}=\int P(\xi)\ln Z=\lim_{m\to 0}\frac{1}{m} \ln \int P(\xi) \;Z^m
\end{equation}
Following standard procedure, we introduce the parameter $Q_{ab}$ by
\begin{equation}
1=\int \prod_{a<b} dQ_{ab}\;\delta(Q_{ab}-\boldsymbol{W}_a\cdot
\boldsymbol{W}_b/N). 
\end{equation}
Hence
\begin{equation*} \int P(\xi) \;Z^m
=\int P(\xi) \;\int \prod_{a=1}^m d\boldsymbol{W}_a\;e^{-\beta H_\xi(\boldsymbol{W}_a)}
=:\int \prod_{a<b}dQ_{ab}\; e^{NA[\{Q_{ab}\}]}.
\end{equation*}
As $N\to\infty$ we compute the stability equations for $A[Q]$ and
obtain $Q$. It is known that the so called replica-symmetry (RS)
ansatz $Q_{ab}=(1-Q)\delta_{ab}+Q$ leads to the correct result for the
spherical perceptron, while it is not correct in the
discrete case.
Hence, the limit $Q\to 1^-$ identifies the point in which all the
solutions collapse onto a single one (up to subextensive
contributions) and it yields the critical point $\alpha_c$~\cite{Nishimori:2001}.

In our model, this procedure (outlined in Appendix \ref{A}) yields
\begin{equation}
\overline{F_{\alpha}(D_{\textrm{out}}|d_{\textrm{in}})}=
\lim_{m\to 0}\frac{1}{Nm}\\
\ln\int dQ \;d\hat Q\; e^{Nm[G_0(\hat Q)+G_1(Q,d_{\textrm{in}},D_{\textrm{out}})-iQ\hat Q/2]}.
\end{equation}
The function $G_0$ depends on the model. For the spherical perceptron
\begin{equation}
G_0^{\textrm {spherical}}(\hat Q)=-\sqrt{i\hat Q}-i\hat Q/2,
\end{equation}
while for the discrete model
\begin{equation}
G_0^{\textrm{discrete}}(\hat Q)=\ln2-i\hat Q/2+\frac{1}{\sqrt{2\pi}}\int dx\; \exp\left(-\frac{1}{2}x^2\right)\ln\cosh\left[x\sqrt{i\hat Q}\right].
\end{equation}
The second function is common to the two models:
\begin{equation}
\label{G1}
\begin{split}
G_1(&Q,d_{\textrm{in}},D_{\textrm{out}})=\\
&(1-D_{\textrm{out}})\left[\int \mathcal{D}_{d_{\textrm{in}}}(y,\bar y)\ln\int_{x>y;\;\bar x>\bar y} \mathcal{D}_{d_{\textrm{in}}}\left(\sqrt{\frac{Q}{1-Q}}x,\sqrt{\frac{Q}{1-Q}}\bar x\right)\right] \\
+&D_{\textrm{out}} \left[\int \mathcal{D}_{d_{\textrm{in}}}(y,\bar y)\ln\int_{x>y;\;\bar x<\bar y} \mathcal{D}_{d_{\textrm{in}}}\left (\sqrt{\frac{Q}{1-Q}}x,\sqrt{\frac{Q}{1-Q}}\bar x\right )\right]\\
+&D_{\textrm{out}}\ln D_{\textrm{out}}+(1-D_{\textrm{out}})\ln(1-D_{\textrm{out}}),
\end{split}
\end{equation}
where
\begin{equation}
\mathcal{D}_{d_{\textrm{in}}}(x,\bar x)=dx\;d\bar x \frac{1}{2\pi}\exp\Bigg(-\frac{1}{2\sqrt{1-q^2}}\begin{bmatrix}x & \bar x\end{bmatrix}\begin{bmatrix}1&-q \\-q &1\end{bmatrix}\begin{bmatrix}x\\ \bar x\end{bmatrix}\Bigg),
\end{equation}
with $q=1-2d_{\textrm{in}}$. It is worth noticing that in the limits
$d_{\textrm{in}}\to 0$ and $D_{\textrm{out}}\to 0$, $G_0$ and $G_1$
reduce to Gardner and Derrida's RS
formulas~\cite{Gardner:JPA:1988}. Finally, the saddle point equations
are
\begin{equation}
-i\hat Q/2=\frac{d}{dQ}G_1\\ \quad \quad
-i Q/2=\frac{d}{d\hat Q}G_0
\end{equation}
and are studied in Appendix \ref{B}.

In order to compute the generalization capacity
$\alpha_c(D_{\textrm{out}}|d_{\textrm{in}})$, we expand both sides of
the equations and match the leading divergent terms. The result is:
\begin{equation}\label{ac(Dd)}
\begin{split}
\alpha_c(D_{\textrm{out}}|d_{\textrm{in}})=\alpha_c
&\left[
(1-D_{\textrm{out}})\left(1+\frac{4}{\pi} \arctan\sqrt{\frac{d_{\textrm{in}}}{1-d_{\textrm{in}}}}\right)\right.\\
&\;\left.+D_{\textrm{out}}\left(1+\frac{4}{\pi} \arctan\sqrt{\frac{1-d_{\textrm{in}}}{d_{\textrm{in}}}}\right)
\right]^{-1}
\end{split}
\end{equation}
with $\alpha_c=\alpha_c(0|0)$ being the RS capacity. It is $4/\pi$ for
the discrete perceptron (which is known to be incorrect) and
(correctly) $2$ for the spherical perceptron. Hence, the RSB
computation would be needed for the discrete case.

The distance-based capacity
$\alpha_c(D_{\textrm{out}}|d_{\textrm{in}})$ is manifestly symmetric
with respect to
$$ (d_{\textrm{in}},D_{\textrm{out}})\mapsto(1-d_{\textrm{in}},1-D_{\textrm{out}})$$

\subsubsection{Analytical and numerical phase diagram}

Notably, the limit $Q\to 1$ does not commute with those for
$d_{\textrm{in}}\to 0^+$ and $d_{\textrm{in}}\to 1^-$,
where the capacity is discontinuous.
Some notable points are
\begin{equation} \label{l1}\lim_{d_{\textrm{in}}\to 0^+}\alpha_c(1|d_{\textrm{in}})=\frac{\alpha_c}{3}\end{equation}
\begin{equation} \label{l2}\lim_{d_{\textrm{in}}\to 1^-}\alpha_c(0|d_{\textrm{in}})=\frac{\alpha_c}{3}\end{equation}
\begin{equation} \label{l3}\alpha_c(D_{\textrm{out}}\neq
  0|0)=\alpha_c(D_{\textrm{out}}\neq 1|1)= 0. \end{equation} 
These
points show the existence of a critical value $\alpha=\alpha_c/3$
below which any degree of generalization is possible. More
specifically, from (\ref{l1}) and (\ref{l3}), we see that, below this
value, it is typically possible to distinguish arbitrarily similar
input sets, unless they are identical. On the other hand, from
(\ref{l2}) and (\ref{l3}), we see that it is possible to identify
input sets which are arbitrarily different (unless they are exactly
anti-parallel).

Moreover, $\alpha_c(0|d_{\textrm{in}})$, displays a vertical tangent
at $(0,\alpha_c)$. This means that, when the size of two input sets is
close to the typical threshold value $\alpha_c N$, then their distance
$\Omega$ typically goes to $\infty$ even for small values of their
distance $d_{\textrm{in}}$. In this limit, it is typically impossible
for a perceptron  to identify (give similar outputs) two training sets even if they are very similar.

\begin{figure}[ht]
\centering
     \includegraphics[width=0.9\textwidth]{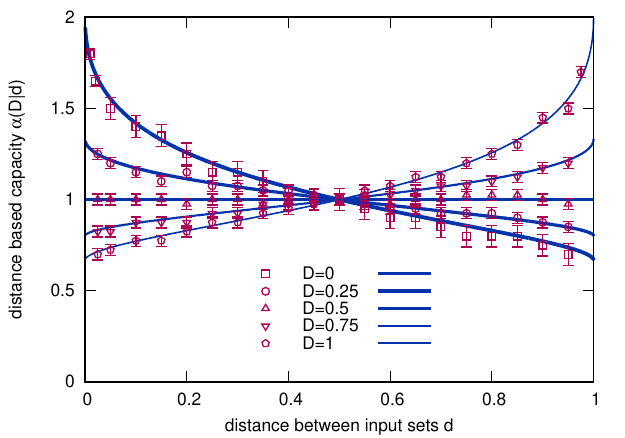}
     \caption{Distance-based capacity of a perceptron.  The
       analytical prediction (solid lines) for the distance-based
       capacity, Eq.~(\ref{ac(Dd)}), is compared to numerical
       calculations (symbols), for the spherical perceptron.  The five
       symbols correspond to the different values of $D_\mathrm{out}$ reported in
       the legend.  The lowest capacity, $\alpha=2/3$, is achieved by
       $\alpha_\mathrm{c}(D_\mathrm{out}=0|d_\mathrm{in}=1)$ and
       $\alpha_\mathrm{c}(D_\mathrm{out}=1|d_\mathrm{in}=0)$.  }
\label{figure:sim}
\end{figure}

Fig.~\ref{figure:sim} shows the agreement between the analytical
formula (\ref{ac(Dd)}) and numerical calculations, in the case of the 
spherical perceptron.

\section{Discussion and Conclusions}

The distance-based approach to generalization proposed here is
complementary to existing ones, and quantifies how a network can
recognize input sets having a prescribed degree of similarity
to the training set.  This approach corresponds to treating
generalization as a task-dependent feature, rather than an
absolute property of a neural network.
In this framework, a network generalizes well if, given
a training set, a typical network configuration assigns similar
outputs to input patterns with fixed dissimilarity.

In order to demonstrate the applicability of our approach, we have
performed explicit calculations for the perceptron, considering
uncorrelated random inputs, with the standard Hamming distance. Our
calculations lead to the distance-based generalization capacity
$\alpha_c(D_\mathrm{out}|d_\mathrm{in})$.  The resulting phase diagram
indicates that generalization is possible at any distance, but with
decreasing critical capacity.  The critical capacity has steeper drops
close to the minimal and maximal distances, while showing slower
decrease for intermediate increasing distances between the sets of
patterns.
The formula we have obtained for the critical line,
Eq.~(\ref{ac(Dd)}), is formally equivalent to the one computed in a
classic study~\cite{Lopez:JPA:1995} with completely different
motivations.  The special case $D_\mathrm{out}=0$ was very recently
rediscovered in a completely different
context~\cite{Chung:PRE:2016,Chung:PRX:2018}, where it represents the
capacity of a perceptron trained to discriminate segments.
We surmise that the statistical-mechanics literature on neural
networks, spanning the last 40 years, is replete with technical
results awaiting to be rediscovered and reinterpreted in more
contemporary settings.

The concept of distance-based capacity is general and may be useful in
numerical experiments with several kinds of neural networks.  Rigorous
and computationally feasible definitions of distances for real-world
objects are important and challenging, as is well-known, for example,
for images~\cite{Gesu:PRL:1999}.
We have chosen here a simple instance of pattern distance but we
expect that applying our approach using different system-tailored
choices of distance metrics could reveal many aspects of how even
complex neural networks operate.
The distance-based approach may also be useful for gaining a better
understanding of the phenomenon of catastrophic
forgetting~\cite{McCloskey:book}, i.e., the quick increase in the
number of errors done on a first training set after the network is
trained on a second set.


Finally, we comment briefly about a potential application of
$F(D_{\textrm{out}}|d_{\textrm{in}})$ to probe the landscape of
synaptic solutions.
The existence of difference classes of solutions has been established
in the case of the discrete simple perceptron. While dominant
solutions are isolated, there exist clustered subdominant solutions,
within a certain
range~\cite{Huang:PRE:2014,Baldassi:PRL:2015,Baldassi:PNAS:2016}.
%
Suppose the numbers $N_\mathrm{d}$ and $N_\mathrm{s}$ of dominant and
subdominant solutions are exponentially large in the size $N$, with
different rates $\Sigma_\mathrm{d}$ and $\Sigma_\mathrm{s}$ such that
$\Sigma_\mathrm{d}>\Sigma_\mathrm{s}$.
Then the subdominant class has zero measure in the thermodynamic
limit, and is therefore ``invisible'' to conventional statistical
averaging.
Now consider the $F(0|d)$ solutions to the problem of learning two
input sets $\xi$ and $\bar\xi$ at distance $d_{\textrm{in}}$.
$\Sigma_\mathrm{d}$ and $\Sigma_\mathrm{s}$, as functions of
$d_{\textrm{in}}$, could cross each other at some value
$d_{\textrm{in}}=\tilde d_{\textrm{in}}$, so that the subdominant set
becomes dominant beyond $\tilde d_{\textrm{in}}$.  This would be
signaled by a discontinuity in $F(0|d_{\textrm{in}})$:
\begin{equation}\label{conjecture}
F(0|d_{\textrm{in}})=\lim_{N\to\infty} \frac{1}{N}\ln\left[\exp N\Sigma_\mathrm{d}(d_{\textrm{in}})+\exp N\Sigma_\mathrm{s}(d_{\textrm{in}})\right]
=\max[\Sigma_\mathrm{d}(d_{\textrm{in}}),\Sigma_\mathrm{s}(d_{\textrm{in}})].
\end{equation}
We speculate that this may be the case if dominant solutions are
isolated and more sensitive to small input differences than clustered
ones.  More in general, non-analytic points of
$F(D_{\textrm{out}}|d_{\textrm{in}})$ may reveal a non trivial
landscape of solutions, highlighting the presence of different
synaptic classes.

\appendix

\section{Replica computation}\label{A}

In this section we will compute $F(D_{\textrm{out}}|d_{\textrm{in}})$ with the replica formalism:
\begin{align}
F(D_{\textrm{out}}|d_{\textrm{in}})&=\lim_{m\to0}\frac{1}{Nm}\ln\int P_{d_{\textrm{in}}}(\xi,\bar\xi)\prod_{a=1}^m\sum_{\{\epsilon_a^\mu=\pm1\}}\prod_{a=1}^m\delta\left(\sum_{\mu=1}^p\epsilon_{\mu}^a-p(1-2D_{\textrm{out}})\right)\nonumber \\
&\int\prod_{a=1}^m dP(W^a)\prod_{\mu=1}^p\prod_{a=1}^m\Theta\left(W^a\cdot \xi_\mu\right)\;\Theta\left(\epsilon_\mu^a\; W^a\cdot \bar\xi_\mu\right)
\end{align}
We can introduce the parameters $Q_{ab}$ and $\hat Q_{ab}$ by inserting the identy
$$ 1=\int \prod_{a<b}dQ_{ab}\delta(Q_{ab}-W_a\cdot W_b/N)=N^m\int \prod_{a<b}dQ_{ab}\;d\hat Q_{ab}\;e^{i N\hat Q_{ab}\;(Q_{ab}-W_a\cdot W_b/N)}$$
in the integral $\int \prod_{a=1}^m dP(W_a)$. With this choice, $F(D_{\textrm{out}}|d_{\textrm{in}})$ can be rewritten as
\begin{equation}\label{equ}
F(D_{\textrm{out}}|d_{\textrm{in}})=\lim_{m\to0}\frac{1}{Nm}\ln \int \prod_{a<b}dQ_{ab}\;d\hat Q_{ab}\;e^{mN\left[i\sum_{a<b}Q_{ab}\hat Q_{ab}+G_0(\{\hat Q_{ab}\})+G_1(\{Q_{ab}\},D_{\textrm{out}},d_{\textrm{in}})  \right]}
\end{equation}
with
\begin{equation}
G_0(\{\hat Q_{ab}\})=\frac{1}{mN}\ln\int \prod_{a=1}^m dP(W_a)\;e^{-i\sum_{a<b}\hat Q_{ab}\; W_a\cdot W_b} 
\end{equation}
and
\begin{align}
G_1(\{ Q_{ab}\},D_{\textrm{out}},d_{\textrm{in}})&=\frac{1}{mN}\ln\int P_{d_{\textrm{in}}}(\xi,\bar\xi)\sum_{\{\epsilon_a^\mu=\pm1\}}\prod_{a=1}^m\delta\left(\sum_{\mu=1}^p\epsilon_{\mu}^a-p(1-2D_{\textrm{out}})\right) \times \nonumber \\ & \times \prod_{\mu=1}^p\prod_{a=1}^m\Theta\left(W^a\cdot \xi_\mu\right)\;\Theta\left(\epsilon_\mu^a\; W^a\cdot \bar\xi_\mu\right)
\end{align}
The function $G_1$ has two properties:
\begin{itemize}
\item it depends on $\{W_a\}$ only though the overlap $W_a\cdot W_b/N= Q_{ab}$ 
\item it can be rewritten in terms of Gaussian varibles $x_a^\mu=W_a\cdot \xi_\mu/\sqrt{N}$.
\end{itemize}
Both these properties can be deduced from the joint distribution of the auxiliary variables $\{x_a^\mu,\bar x_a^\mu\}_{a=1,...,m}$:
\begin{multline}\label{gauss}
P^\mu_{q}(\{x^\mu_a,\bar x^\mu_a\}):= \frac{1}{P(d_{\textrm{in}})}P_{d_{\textrm{in}}}(\xi_\mu,\bar\xi_\mu)\prod_{a=1}^m\delta(x_a^\mu-W_a\cdot \xi^\mu/\sqrt{N})\;\delta(\bar x_a^\mu-W_a\cdot \bar \xi^\mu/\sqrt{N})\\
\underset{N\to\infty}{\to}\frac{1}{(2\pi)^m\sqrt{\det M(Q,q)}}\exp\left( -\frac{1}{2}\begin{bmatrix}x^\mu\\ \bar x^\mu\end{bmatrix}^t  M^{-1}(Q,q)  \begin{bmatrix}x^\mu\\ \bar x^\mu\end{bmatrix}   \right) 
\end{multline}
with
\begin{equation}
x=\begin{bmatrix}x_1^\mu\\...\\  x_m^\mu\end{bmatrix}\hspace{1cm} \bar x=\begin{bmatrix}\bar x_1^\mu\\...\\  \bar x_m^\mu\end{bmatrix}\hspace{1cm}M(Q,q)=\begin{bmatrix}Q &qQ\\ qQ & Q\end{bmatrix}\hspace{1cm} q=1-2d_{\textrm{in}}
\end{equation}
and $Q$ being a shorthand notation for the matrix $Q_{ab}$. (\ref{gauss}) only holds in the thermodynamic limit $N\to\infty$: a key passage consists in expanding $\ln\cosh(y/\sqrt{N})\sim -\frac{y^2}{2N}$ for some finte $y$. Therefore
\begin{equation}\label{g1pre}
G_1(\{ Q_{ab}\},D_{\textrm{out}},d_{\textrm{in}})=\frac{1}{N}\ln \sum_{\{\epsilon_a^\mu=\pm1\}}\prod_{a=1}^m\delta\left(\sum_{\mu=1}^p\epsilon_{\mu}^a-p(1-2D_{\textrm{out}})\right)\prod_{\mu=1}^p A_\mu(\{\epsilon^\mu_a\},Q,q)
\end{equation}
with
\begin{equation}\label{Apre}
A_\mu(\{\epsilon^\mu_a\},Q,q)= \int \prod_{a=1}^m dx_a^\mu\;d\bar x_a^\mu\;P^\mu_{q}(\{x^\mu_a,\bar x^\mu_a\},Q) \;\prod_{a=1}^m\Theta(x_\mu^a)\;\Theta(\epsilon_a^\mu\;\bar x_\mu^a).
\end{equation}

We can now introduce the RS ansatz 
\begin{equation}
\label{rsansatz}Q_{ab}=Q+(1-Q)\delta_{ab}.
\end{equation}
Under this assumption, we can compute $G_0$ and $G_1$. The computation of $G_0$ is straightforward. In the discrete case,
\begin{equation}
G_0=\ln \cbk{2^me^{-im\hat Q/2}\frac{1}{\sqrt{2\pi i\hat Q}}\int dx\; \exp\rbk{-\frac{1}{2i\hat Q}x^2+m\ln\cosh(x)}}.
\end{equation}
If we expand it for small $m$, we get
\begin{equation}
G_0= m\cbk{\ln2-i\hat Q/2+\frac{1}{\sqrt{2\pi}}\int dx\; \exp\rbk{-\frac{1}{2}x^2}\ln\cosh\rbk{x\sqrt{i\hat Q}}} +o(m).
\end{equation}
In the spherical case,
\begin{equation} G_0=-imJ-(1/2)\ln\rbk{\frac{iJ+(m-1)i\hat Q/2}{(iJ-i\hat Q/2)^{1-m}}}\end{equation}
with
$$ iJ=\frac{1}{2}(\sqrt{i\hat Q}+i\hat Q).$$
For small $m$
\begin{equation}
G_{0}(\hat Q)=-m\cbk{\sqrt{i\hat Q}+i\hat Q/2}.
\end{equation}

We can compute $G_1$ from (\ref{g1pre}). After some formal manipulations, the functions $A_\mu$, as given by (\ref{Apre}), can be rewritten as
\begin{multline}
A_\mu(D_\mu,Q,q)=
\int \mathcal{D}_d(y,\bar y)\Bigg[\int_{x>y;\;\bar x<\bar y} \mathcal{D}_d\left (\sqrt{\frac{Q}{1-Q}}x,\sqrt{\frac{Q}{1-Q}}\bar x\right )\Bigg]^{mD_\mu} \times \\
\times\Bigg[\int_{x>y;\;\bar x>\bar y} \mathcal{D}_d\left (\sqrt{\frac{Q}{1-Q}}x,\sqrt{\frac{Q}{1-Q}}\bar x\right )\Bigg]^{m(1-D_\mu)}
\end{multline}
with
\begin{equation}
\mathcal{D}_d(x,\bar x)=dx\;d\bar x \frac{1}{2\pi}\exp\Bigg(-\frac{1}{2\sqrt{1-q^2}}\begin{bmatrix}x & \bar x\end{bmatrix}\begin{bmatrix}1&-q \\-q &1\end{bmatrix}\begin{bmatrix}x\\ \bar x\end{bmatrix}\Bigg)
\end{equation}
and
\begin{equation}
m(1-2D_\mu)=\sum_{a=1}^m \epsilon_\mu^a.
\end{equation}
For small $m$
\begin{multline}\label{A2}
A_\mu(D_\mu,Q,q)=
\int \mathcal{D}_d(y,\bar y)\Bigg[mD_\mu\ln\int_{x>y;\;\bar x<\bar y} \mathcal{D}_d\left (\sqrt{\frac{Q}{1-Q}}x,\sqrt{\frac{Q}{1-Q}}\bar x\right ) \\ +m(1-D_\mu)\ln
\int_{x>y;\;\bar x>\bar y} \mathcal{D}_d\left (\sqrt{\frac{Q}{1-Q}}x,\sqrt{\frac{Q}{1-Q}}\bar x\right )\Bigg]
\end{multline}
up to $o(m)$. Now, let us observe that
\begin{equation} \label{dpm}\sum_{\mu=1}^p mD_\mu=\frac{mp}{2}-\frac{1}{2}\sum_{\mu=1}^p\sum_{a=1}^m \epsilon_\mu^a=D_{\textrm{out}}\, p\, m .
\end{equation}
If we combine (\ref{dpm}), (\ref{A2}) and (\ref{g1pre}), we get (\ref{G1}) 
\begin{multline}\label{k}
m\alpha G_1(Q,d,D_{\textrm{out}})=m\alpha(1-D_{\textrm{out}})\left[\int \mathcal{D}_d(y,\bar y)\ln\int_{x>y;\;\bar x>\bar y} \mathcal{D}_d\left(\sqrt{\frac{Q}{1-Q}}x,\sqrt{\frac{Q}{1-Q}}\bar x\right)\right]\\
 +
m\alpha D_{\textrm{out}} \left[\int \mathcal{D}_d(y,\bar y)\ln\int_{x>y;\;\bar x<\bar y} \mathcal{D}_d\left (\sqrt{\frac{Q}{1-Q}}x,\sqrt{\frac{Q}{1-Q}}\bar x\right )\right]\\
+m\alpha[D_{\textrm{out}}\ln D_{\textrm{out}}+(1-D_{\textrm{out}})\ln(1-D_{\textrm{out}})]+o(m).
\end{multline}
For the sake of simplicity, we can rewrite (\ref{k}) as
\begin{align}\label{G1A}
G_1(Q,d,D_{\textrm{out}})&=:(1-D_{\textrm{out}})A_+(Q,d,D_{\textrm{out}}) + D_{\textrm{out}} A_-(Q,d,D_{\textrm{out}}) + \nonumber  \\&+D_{\textrm{out}}\ln D_{\textrm{out}}+(1-D_{\textrm{out}})\ln(1-D_{\textrm{out}}).
\end{align}

\section{Saddle point equations and derivation of the critical capacity}\label{B}
From (\ref{equ}) we read the saddle point equations: 
\begin{align}
&iQ_{ab}=\frac{d}{d\hat Q_{ab}} \label{sp1}G_0\\
&i\hat Q_{ab}=\frac{d}{d Q_{ab}}\label{sp2}G_1.
\end{align}
If we use the RS ansatz (\ref{rsansatz}), then (\ref{sp1}) and (\ref{sp2}) become
\begin{align}
&-imQ/2=\frac{d}{d\hat Q} \label{RSsp1}G_0\\
&-im\hat Q/2=\frac{d}{d Q}\label{RSsp2}G_1
\end{align}
as $m\to 0$. 
Eq.~(\ref{RSsp2})  can be rewritten as (see (\ref{G1A}) for the notation)
\begin{flalign} \label{eq1}
   -\frac{i}{2}\hat Q&=\alpha\frac{d}{dQ}[DA_-+(1-D)A_+].
\end{flalign}
Eq.~(\ref{RSsp1}) for the discrete model is
\begin{flalign}
-\frac{1}{2} Q&=-1/2+\frac{1}{\sqrt{8\pi i\hat Q}}\int dx\; x\, \exp\rbk{-\frac{1}{2}x^2}\tanh\rbk{x\sqrt{i\hat Q}}   \label{lna}         
\end{flalign}
while for the spherical model
\begin{equation}
-\frac{1}{2}Q=-\frac{1}{2}\rbk{1+1/\sqrt{i\hat Q}}.
\end{equation}
From the previous equations, we can write
\begin{equation}
\alpha(d_{\textrm{in}},D_{\textrm{out}},Q)=\frac{-i\hat Q(Q)}{2\frac{d}{d Q}[D_{\textrm{out}}A_-(d_{\textrm{in}},Q)+(1-D_{\textrm{out}})A_+(d_{\textrm{in}},Q)]}.
\end{equation}
In order to get the critcal capacity, we should evaluate $\alpha(d_{\textrm{in}},D_{\textrm{out}},Q)$ at $Q=1$:
\begin{equation}\label{ac0}
\alpha_c(D_{\textrm{out}}|d_{\textrm{in}})=\lim_{Q\to1^-}\frac{-i\hat Q(Q)}{2\frac{d}{dQ}[D_{\textrm{out}}A_-(d_{\textrm{in}},Q)+(1-D_{\textrm{out}})A_+(d_{\textrm{in}},Q)]}.
\end{equation}
 However, the quantities appearing in the previous equation are neither defined nor analytical at $Q=1$. For this reason we have to extract the leading divergencies around $Q=1^-$. We have to expand $D_{\textrm{out}}A_-(d_{\textrm{in}},Q)+(1-D_{\textrm{out}})A_+(d_{\textrm{in}},Q)$ first. For this purpose, it is convenient to perform a change of variables $S=(y+\bar y)/\sqrt 2$, $D=(y-\bar y)/\sqrt 2$ (not to be confused with the output difference), $s=(x+\bar x)/\sqrt 2$ and write:
\begin{multline} \label{a-}
 A_{-}=\frac{1}{2\pi}\int dS\;dD\;\exp\rbk{-\frac{1}{2}S^2-\frac{1}{2}D^2}
\ln \frac{1}{\sqrt{2\pi}} \times \\ \times \Bigg\{\int_{0}^{\infty}ds\; \exp\rbk{-\frac{1}{2}\frac{Q}{1-Q} (s-S)^2}\erfc\rbk{\sbk{\sqrt{\frac{Q}{1-Q}}\sbk{D+\sqrt{\frac{1+q}{1-q}}s} }}\\
+\int_{0}^{\infty}ds\; \exp\rbk{-\frac{1}{2}\frac{Q}{1-Q} (s+S)^2}
\erfc\rbk{\sbk{\sqrt{\frac{Q}{1-Q}}\sbk{D+\sqrt{\frac{1+q}{1-q}}s} }}\Bigg\}\\
\end{multline}
\begin{multline}\label{a+}
A^+=\frac{1}{2\pi}\int dS \;dD\; \exp\rbk{-\frac{1}{2}S^2-\frac{1}{2}D^2}
\ln \frac{1}{\sqrt{2\pi}}\sqrt{\frac{1-Q}{Q}}\int_{0}^\infty ds\;\exp\rbk{-\frac{1}{2}\,\frac{Q}{1-Q}\,(s+S)^2}\\
\Bigg\{\erfc\rbk{\sqrt{\frac{Q}{1-Q}}\sbk{D-\sqrt{\frac{1+q}{1-q}}s} }
-      \erfc\rbk{\sqrt{\frac{Q}{1-Q}}\sbk{ D+\sqrt{\frac{1+q}{1-q}}s} }\Bigg\}
\end{multline}
where we have used the definition $\erfc(y)=\frac{1}{\sqrt{2\pi}}\int_y^\infty e^{-x^2/2}$. We can use the following limit in order to perform the expansion:
\begin{equation}
\lim_{\epsilon\to 0^+}\epsilon\ln \int_I dx\; e^{-a x^2/\epsilon}\erfc((b x-c)/\sqrt{\epsilon})=\min_{x\in I}(a x^2+\Theta(bx-c)(bx-c)^2).
\end{equation}
If we set $\epsilon=1-Q$ in (\ref{a-}) and (\ref{a+}), then we obtain that
\begin{equation}
A_\pm(Q\to1)\sim\frac{1}{1-Q}\int \mathcal{D}x\;\mathcal{D}y\; \sum_j P_j^\pm(x,y)\;\chi_{\Omega_j},
\end{equation}
where $\mathcal{D}x=dx\;\frac{1}{\sqrt{2\pi}}e^{-x^2/2}$, $\chi$ is the characteristic function, $P_j$s are second degree polynomials and $\Omega_j$s are disjoint sets such that $\cup_j \Omega_j=\mathbb{R}^2$. The explicit computation is a bit lenghty but straightforward and the result is
\begin{equation}\label{Apm}
A_\pm(Q\to 1)\sim\frac{1}{1-Q}\cbk{1+\frac{4}{\pi}\arctan\sbk{\rbk{\frac{1-q}{1+q}}^{\pm1/2}}}.
\end{equation}
However, while $A_\pm(Q\to 1)$ (and their derivatives) are finite at both $d_{\textrm{in}}=1$ and $d_{\textrm{in}}=0$, if we fix $Q$, then
\begin{equation}
\lim_{d_{\textrm{in}}\to 0^+} A_-=\lim_{d_{\textrm{in}}\to 1^-} A_+=-\infty
\end{equation}
since
$$ \lim_{\epsilon\to 0^+}\int dx\;e^{-ax^2}\; \erfc(x/\epsilon-b)=0.$$
Therefore
\begin{equation}\label{lim1}\lim_{d_{\textrm{in}}\to1^-}\frac{-i\hat Q(Q)}{2\frac{d}{dQ}[D_{\textrm{out}}A_-(d_{\textrm{in}},Q)+(1-D_{\textrm{out}})A_+(d_{\textrm{in}},Q)]}=0\end{equation}
unless $D_{\textrm{out}}=1$, and
\begin{equation}\label{lim2}\lim_{d_{\textrm{in}}\to0^+}\frac{-i\hat Q(Q)}{2\frac{d}{dQ}[D_{\textrm{out}}A_-(d_{\textrm{in}},Q)+(1-D_{\textrm{out}})A_+(d_{\textrm{in}},Q)]}=0\end{equation}
unless $D_{\textrm{out}}=0$. The conclusion is that the limits in $(D_{\textrm{out}},d_{\textrm{in}})$ and $Q$ do not commute. The consequence is the existence of a delta discontinuity. Finally, we can combine (\ref{lim1}), (\ref{lim2}), (\ref{Apm}) and (\ref{ac0}) into the distance-based capacity in (\ref{ac(Dd)}).

\bibliographystyle{unsrt}
\bibliography{paperbib}

\end{document}